\documentclass[preprint,superscriptaddress,amsmath,amssymb]
{elsarticle}
\usepackage{graphicx} 
\usepackage[export]{adjustbox}
\usepackage{dcolumn}
\usepackage{bm}
\usepackage[english]{babel}
\usepackage{amsmath,amssymb}
\usepackage[utf8]{inputenc}
\usepackage{psfrag}
\usepackage{xcolor}
\usepackage{hyperref}
\usepackage{appendix}
\usepackage[normalem]{ulem}
\usepackage{lipsum}
\usepackage{tikz}
\usepackage{mathtools}

\begin{document}

\title{The Evolution of Pluralistic Ignorance}

\author[1,2]{Alessandra F. Lütz\corref{cor1}}
\ead{sandiflutz@gmail.com}

\author[2]{Lucas Wardil}
\ead{lucaswardil@gmail.com}

\cortext[cor1]{Corresponding author}

\affiliation[1]{organization={Instituto de Humanidades, Artes e Ciências, Universidade Federal do Sul da Bahia},
postcode={45988-05},
city={Teixeira de Freitas - BA},
country={Brazil}}

\affiliation[2]{organization={Departamento de Física, Universidade de Minas Gerais},
postcode={31270-901},
city={Belo Horizonte - MG},
country={Brazil}}

\date{March 2024}

\begin{abstract}
Pluralistic ignorance is a social-psychological phenomenon that occurs when individuals privately hold beliefs that differ from perceived group norms. Traditional models, based on opinion dynamics with private and public states, fail to account for a key aspect: when nonexpression aligns with normative behavior, initial social pressure can induce pluralistic ignorance. We show that pluralistic ignorance persists under infrequent imitation and strong initial minority influence. Although individuals can overcome this ignorance by the end of interactions, it reemerges in subsequent meetings. However, excessive imitation erases pluralistic ignorance, leading to a uniform state in which internal and external states align. Furthermore, incorporating memory into the internalization process shows that pluralistic ignorance peaks at moderate imitation levels.
\end{abstract}

\begin{graphicalabstract}
\includegraphics[scale=0.8]{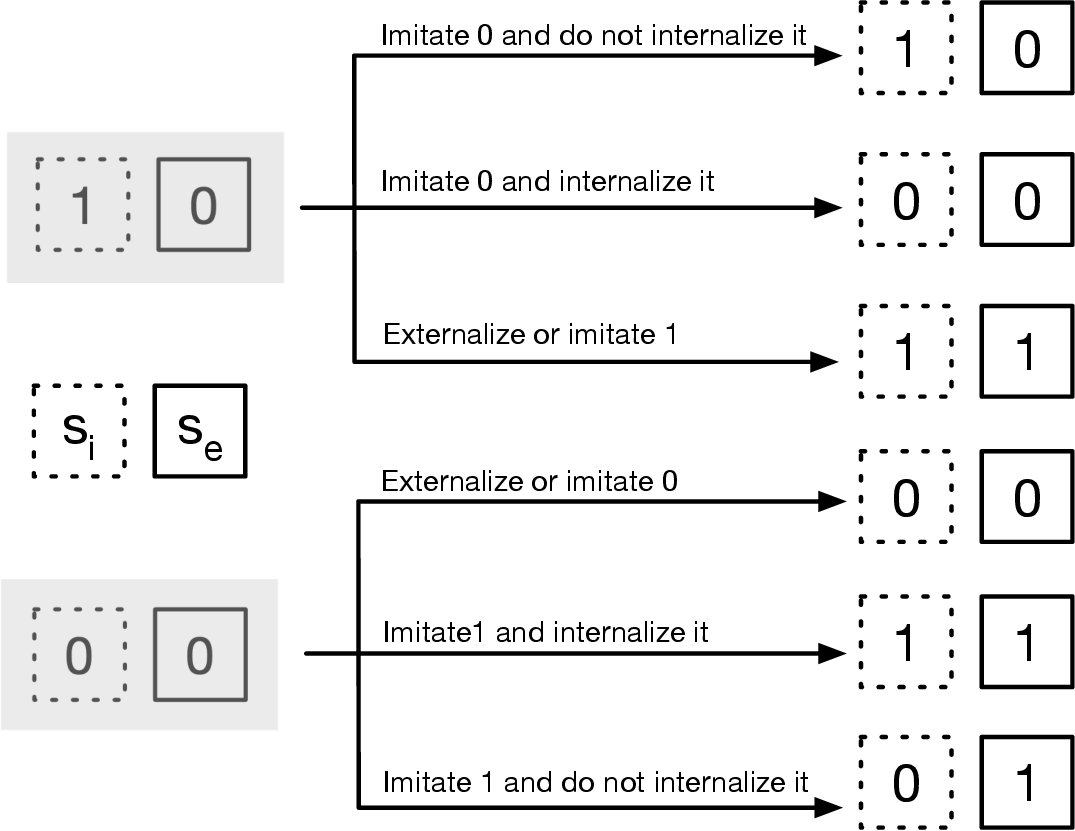}
\end{graphicalabstract}
\begin{highlights}
\item Pluralistic ignorance can emerge from social pressure exerted by the initial conditions.
\item Moderate imitation and strong initial minority promote pluralistic ignorance.
\item Analysis of the role of the first person to break pluralistic ignorance.
\item With memory in internalization, pluralistic ignorance peaks at moderate imitation rates.
\end{highlights}


\maketitle

\section{\label{introduction}Introduction}

Pluralistic ignorance describes a social phenomenon in which individuals privately disagree with a group's perceived norms, but mistakenly believe that their views are uncommon~\cite{sargent2021pluralistic}. This often stems from a desire for social acceptance, leading individuals to conform outward while hiding their true beliefs~\cite{o1975pluralistic,kauffman1981prison,prentice1993pluralistic-alcohol,lambert2003pluralistic}. Pluralistic ignorance can have important implications for group dynamics, decision-making, and social behavior, as it can lead to the perpetuation of norms and beliefs that might not actually reflect the majority's true sentiments. It can also hinder open and honest communication within a group or society~\cite{bicchieri1999great,merdes2017growing}.
In the sociopsychological literature, two classic examples illustrate this phenomenon. The first, the classroom scenario, involves students refraining from asking questions after a challenging lecture, mistakenly believing that they are the only ones confused~\cite{miller1987pluralistic,miller1991social}. The second, the college drinking scenario, describes first-year students who engage in excessive drinking at a party despite finding it unpleasant, influenced by perceived social norms ~\cite{prentice1993pluralistic-alcohol,borsari2001peer}. 
 
The kind of normative behavior in each of these two examples is different due to the contrast between injunctive and descriptive norms~\cite{Cialdini1991AFT}. Injunctive norms relate to the expectations of group conduct imposed by its members, while descriptive norms relate to the behaviors that are prevalently observed within the group. The classroom situation is better categorized under descriptive norms, as students, observing the absence of inquiries from their peers, deduce that such inaction will shield them from appearing incompetent. This initial passivity within the group is subsequently construed as a social norm. In contrast, the drinking scenario exemplifies an injunctive norm, where the dominant normative belief posits that alcohol consumption is viewed positively, seen as an important aspect of the liberal culture of university campuses~\cite{prentice1993pluralistic-alcohol}. Our research delves into these distinctions. Before we advance, let us first review how pluralistic ignorance has been modeled in the specialized literature.


Pluralistic ignorance has been studied since the early 20th century~\cite{bjerring2014rationality-PI}. Despite this long-standing interest, there is still a notable absence of specialized mathematical frameworks specifically designed to elucidate this social phenomenon. The existing models are based on opinion modeling frameworks, which have attracted substantial attention in several scientific disciplines such as sociology, physics, and political science~\cite{castellano2009statistical,xia2011opinion}. Various types of opinion models have emerged, from the Sznajd opinion models~\cite{sznajd2021review}, which investigate peer influence and consensus formation, to Axelrod's cultural dissemination model~\cite{axelrod1997dissemination}, which explores the diffusion of cultural traits. Furthermore, the classic voter model~\cite{clifford1973votermodel}, in which individuals imitate the most popular opinion among their immediate neighbors, is widely used as a framework to study opinion dynamics and social influence. 

Several variations of voter-like models have been proposed~\cite{mobilia2007role,castellano2009nonlinear,fernandez2014voter,gastner2018consensus}. However, because pluralistic ignorance involves a dissonance between people's public actions and their true beliefs or feelings, the models that are more useful to analyze the pluralistic ignorance phenomena are those that consider hidden states. An opinion model that considers public and private opinions is the partisan voter model~\cite{masuda2010heterogeneous,masuda2011can}, in which agents have innate fixed preferences that influence their expressed opinions. A different approach, where hidden opinions may be affected by the expressed ones, is the voter model with hidden opinions~\cite{gastner2018consensus}. In~\cite{gastner2019impact}, an extension of this model is used to study hypocrites, that is, individuals whose internal and external opinions differ. In this model, the mismatch between internal and external opinions arises because there are two processes that occur in two layers: internal and external opinions. However, the mismatch is not caused by a misinterpretation of a social norm, which must be part of the phenomenon of pluralistic ignorance.

Pluralistic ignorance was explicitly modeled~\cite{seeme2016pluralistic,seeme2019pluralistic,seeme2021agent} using an agent-based model that adapts the Voter~\cite{clifford1973votermodel} and Friedkin-Johnsen~\cite{friedkin1990social} models. More specifically, in~\cite{seeme2016pluralistic} individuals hold both internal and external opinions and can imitate the majority opinion in their neighborhood depending on whether they are susceptible or stubborn. 
In~\cite{seeme2019pluralistic}, the model takes into account the benefits of balancing the need to conform to the identity of the group and the need to minimize cognitive dissonance~\cite{seeme2019pluralistic,seeme2021agent}. In both models, public opinions can be internalized if agents are susceptible enough to external influences and the pressure to conform is low enough. Pluralistic ignorance arises when the pressure to conform is above a threshold, forcing agents to adopt public opinions that are different from their internal beliefs.  

In~\cite{seeme2016pluralistic,seeme2019pluralistic,seeme2021agent}, similar to traditional opinion models, individuals continuously manifest their external state, making it perpetually accessible for observation. In contrast, in our proposed model, external states are constructed anew each time a group is formed, and these states are intrinsically distinct, transcending mere differential labeling. In~\cite{seeme2016pluralistic,seeme2019pluralistic,seeme2021agent}, pluralistic ignorance is quantified by the proportion of individuals whose internal state diverges from their external state. Although we agree that this is an indispensable criterion, as evidenced in the study of hypocrites~\cite{gastner2019impact}, the inclusion of a reference to a communal norm is imperative in the definition. Our research introduces an alternative methodology to measure pluralistic ignorance.

Here, we model the pluralistic ignorance phenomenon by setting a pair of internal and external states for each individual. Individuals are sampled from a population to form a group in which interactions take place. As soon as the group is formed, everyone's initial behavior is that of inaction. Thus, the initial external state of all individuals is the same: the one representing inaction. The classroom example demonstrates this key aspect of the model. Although some students may want to ask questions, their initial hesitation implies a collective norm of silence. This perceived norm prevents members from moving out of inaction, reinforcing it as the norm of the group. As the interaction progresses and individuals begin to manifest their internal states externally, it may prompt others within the group to imitate the externalized behavior of others, thereby allowing the actions of a minority to potentially sway the collective decisions in a sequential manner. Individuals may internalize their external state and the composition of the population can evolve. Given that the initial inaction of all individuals in the group is interpreted as normative behavior, those who have an internal state corresponding to an action state (for example, willingness to ask a question) are said to start the group interaction under pluralistic ignorance. We show that pluralistic ignorance emerges even for small imitation probabilities and ceases if the imitation probability is high. However, at the same time, many individuals are able to overcome pluralistic ignorance, with a few remaining with an internal state incompatible with the external one. Furthermore, this research explores the role of memory in the internalization mechanism by enabling individuals to adopt the most frequently exhibited external state from recent interactions.

The paper is structured as follows. The model is defined in the next section~\footnote{The code we use to run our simulations can be found at https://github.com/sandiflutz/pluralistic\_ignorance\_voter .}. The results are presented in three subsections: the first examines random group selection from a memoryless population, the second incorporates memory, and the third models individuals on a square lattice to study spatial correlations. The paper concludes with a general discussion of our findings.

\section{Model}

We consider a population of size $N$ where each individual has an internal state $s_i$ and an external state $s_e$. The inaction state is represented by $0$ and the action by $1$. Thus, each individual is characterized at a given time by the vector $(s_i,s_e)$. For example, individuals of type $(0,1)$ or $(1,0)$ externalize a state different from their internal one. The external state represents the displayed behavior. For example, the external state $1$ can represent the action of asking a question, and the $0$ not asking (notice that here inaction is considered a behavior). The internal state represents the attitude, belief, or inclination toward the action. For example, the internal state $1$ represents the inclination to ask a question, and the $0$ represents indifference or lack of inclination to inquire.

The dynamics involve a stochastic process in which $G$ agents are randomly selected to form a group. Initially, all individuals in the group display the same external state of inaction, that is, the external states of all are set to $0$. Then, one individual at a time (in random order) is chosen to update the external state. This individual can imitate the external state of others in the group, with probability $p_{cp}$, or can externalize the internal state, with probability $1-p_{cp}$. If externalization takes place, the external state is set to the same value as the internal one. If imitation takes place, then with probability $p_1$, given by
\[
p_1=\left(\frac{n_1}{G}\right) ^{1-\gamma}\label{eq.p1},
\]
the individual adopts the external state $1$. With probability $1-p_1$, the individual maintains the external state equal to $0$. Here, $n_1$ is the current number of individuals in the group that have adopted an external state equal to $1$ and $\gamma$ is a parameter controlling the influence of the initial minority ($0<\gamma\le 1$). If $\gamma$ is high, then a single individual externalizing $1$ can trigger the adoption of the state $1$ by others.  Notice that $n_1$ starts equal to zero and can increase as more individuals in the group externalize the state $1$. The first appearance of the external state $1$ in the group is only possible if an individual with an internal state equal to $1$ externalizes it. 

\begin{figure}[h]
\vspace{3mm}
\includegraphics[width=\linewidth]{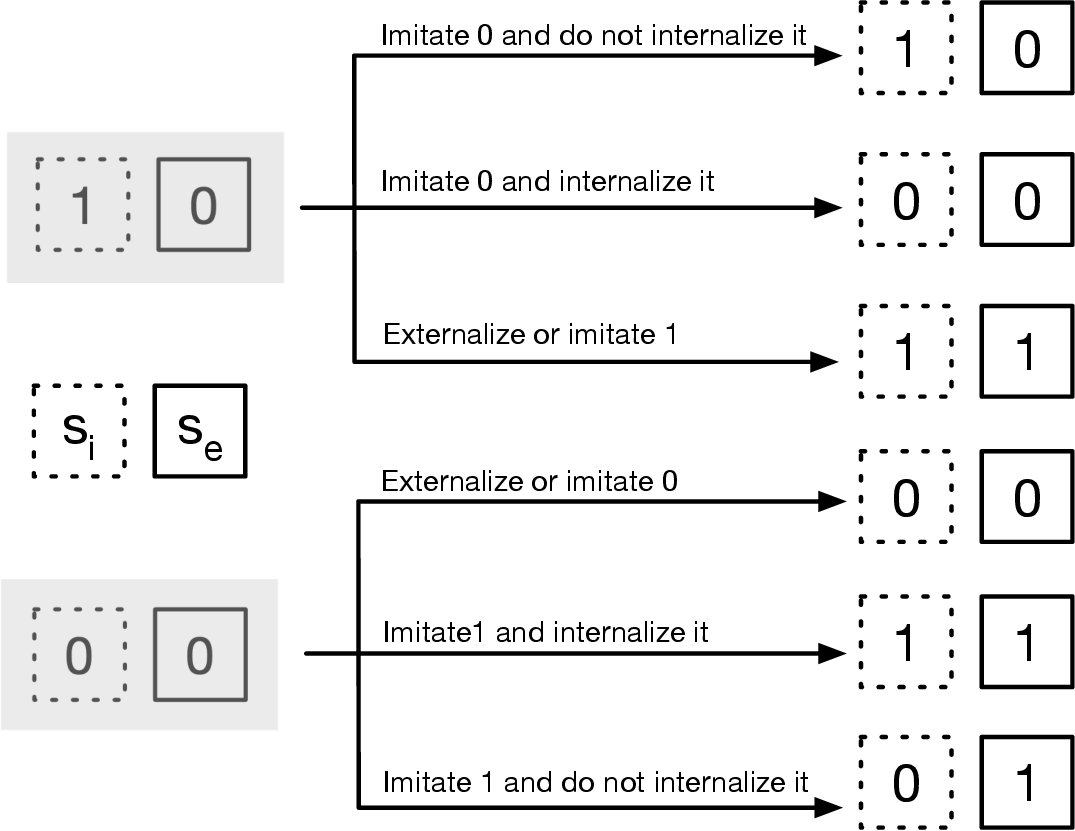}
\caption{Graphical representation of the possible state modifications during a group interaction of memoryless agents ($T_M=1$). The state of an individual is represented by the vector $(s_i,s_e)$, where $s_i$ is the internal state and $s_e$ is the external state. At the beginning of the group interaction, the external states of the individuals are set to zero and they are in one of the following states: $(0,0)$ or $(1,0)$ (grey shadowed rectangles on the left side). Due to imitation, externalization, and internalization, they can finish the group interaction in different states (shown on the right side). For example,  individuals in the $(0,0)$ state can change to $(1,1)$ by imitating the state $1$ and then internalizing it (at the beginning of the next group interaction, the external state is set to zero and this individual starts in the state $(1,0)$).  The pluralistic ignorance level is measured as the fraction of individuals with $s_i=1$, which is the same as the fraction of individuals in the state $(1,0)$ before group interactions take place (states shown on the left side). We also measured the fraction of $(0,0)$, $(0,1)$, $(1,0)$, and $(1,1)$ after all group interactions take place (states shown on the right side).}
\label{fig1}
\end{figure}

After all individuals in the group have the opportunity to display the external state, either by externalization or imitation, each individual can internalize the external state with probability $p_{int}$. The external state that is displayed more often is the one that is internalized. Each individual has a memory of size $T_M$ where they store the last $T_M$ external states that they have displayed. Thus, if internalization takes place, the internal state is set to the same value as the most frequent external state in the memory. For example, if $T_M=1$ the external state adopted in the most recent group interaction is the one that can be immediately internalized. This scenario, where $T_M=1$, is referred to as the case of no memory. Figure~\ref{fig1} illustrates the possible changes in $(s_i,s_e)$ in the update process of an individual. 

The evolution of the population is simulated as follows. First, we initialize half of the individuals with an internal state equal to $1$, and the remaining half to $0$. Then, we sample $G$ individuals to participate in a group interaction in which imitation, externalization, and internalization take place. We repeat this step $N/G$ times so that each individual has the opportunity to participate on average in at least one group. These $N/G$ repetitions comprise one Monte Carlo step (MCS) of the simulation. 

We measure the fraction of each type of internal state in the population in the stationary regime, namely $\rho_{int}^{(0)}$ and $\rho_{int}^{(1)}$  (recall that $\rho_{int}^{(0)}+\rho_{int}^{(1)}=1$). The level of pluralistic ignorance is given by the fraction of individuals with $s_i=1$ in the population, that is, by $\rho_{int}^{(1)}$.  Recall that because, at the beginning of group interactions, everyone displays $s_e=0$, the individuals with $s_i=1$ are those who have the impression that their beliefs differ from those of everyone else. Individuals with $s_i=0$ have no such impression because their internal state matches the behavior observed at the beginning of the interactions.  

We also measure the stationary fraction of individuals in each one of the configurations $(0,0)$, $(0,1)$, $(1,0)$, and $(1,1)$, which can be counted at the end of group interactions. Notice that the external state is the value that is displayed in the most recent participation in a group. Interestingly, the individuals who, after all the interactions in the group take place, end up with $s_i=1$ and $s_e=1$ are those who manage to overcome pluralistic ignorance, expressing their internal state despite the initial opposition. 

In all figures presented in the paper, we show stationary values obtained by time averaging after a transient period. Notice that the population state where all are $(0,0)$ is an absorbing state. Strictly speaking, any fraction of states $(1,0)$, $(0,0)$, or $(1,1)$ are meta-stable states that can persist for very long times if the population size is large enough.

\section{Results}

We considered a system with $N=484$ agents and groups of size $G=20$. The results remain the same for different system sizes. Changing the size of the group has an effect similar to changing $\gamma$. More specifically, multiplying $1-\gamma$ by $\alpha$, with $\alpha>1$, is the same as changing $G$ to $G^\alpha$. For the initial configuration of the population, the internal states are randomly initialized to $0$ or $1$ with equal probability. 
 
\begin{figure}[h]
\includegraphics[width=\linewidth]{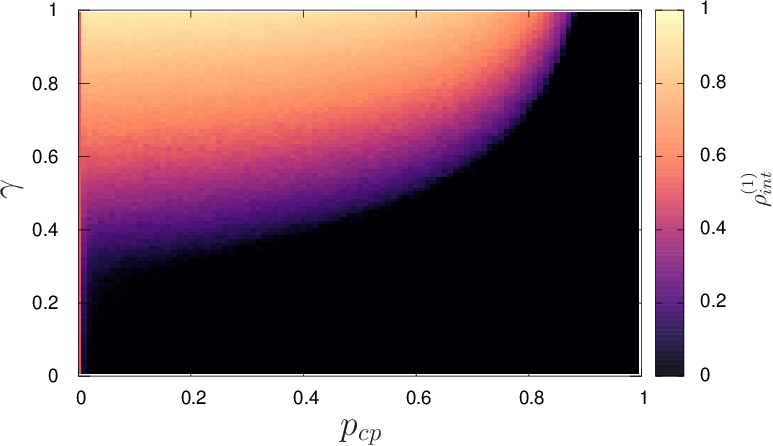}

\caption{Fraction of pluralistic ignorance as a function of the imitation probability ($p_{cp}$) and the influence factor ($\gamma$). The bottom figure shows a cross-section of the top figure at $\gamma=0.9$. Here $p_{int}=0.01$.}
\label{fig2}
\end{figure}

Figure~\ref{fig2} shows the level of pluralistic ignorance for various combinations of imitation probability, $p_{cp}$, and influence factor, $\gamma$. Pluralistic ignorance occurs only when $\gamma$ is high enough and $p_{cp}$ is not too high.
In the absence of imitation ($p_{cp}=0$), the internal states do not change. Alterations occur only after an individual internalizes an imitated state, and this cannot happen for $p_{cp}=0$. Thus, the fraction of individuals with an internal state equal to $1$ and $0$ remains equal to the initial fractions (both of which are set at $0.5$).
If $p_{cp}$ is small, most individuals externalize their internal state. This means that it is very likely that an individual in state $(1,0)$ changes to state $(1,1)$. From now on, if the influence factor $\gamma$ is sufficiently high, it is state 1 that is going to be imitated. Therefore, individuals in the state $(0,0)$ mimic the state $1$ and eventually internalize it, becoming $(1,1)$, increasing the fraction of individuals with an internal state equal to $1$.
If the imitation probability is too high, there are limited opportunities for someone to externalize the state $1$ (recall that with probability $1-p_{cp}$ the individual externalizes the internal state). As a result, most individuals tend to imitate the state $0$, eventually internalizing it. This leads to a final state of homogeneity, where all individuals possess an internal state equal to 0. Notice that, because there is no chance of having an external state equal to $1$, the individuals are trapped in the state $(0,0)$, which is an absorbing state.  

\begin{figure}[h]
\includegraphics[width=0.9\linewidth]{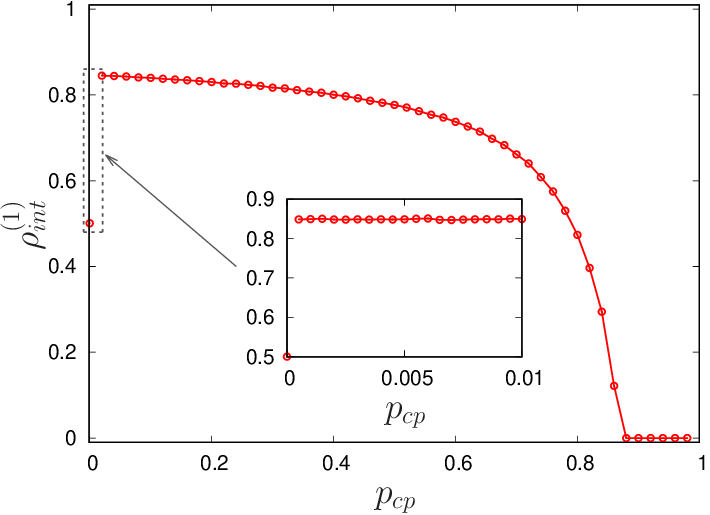}

\includegraphics[width=0.9\linewidth]{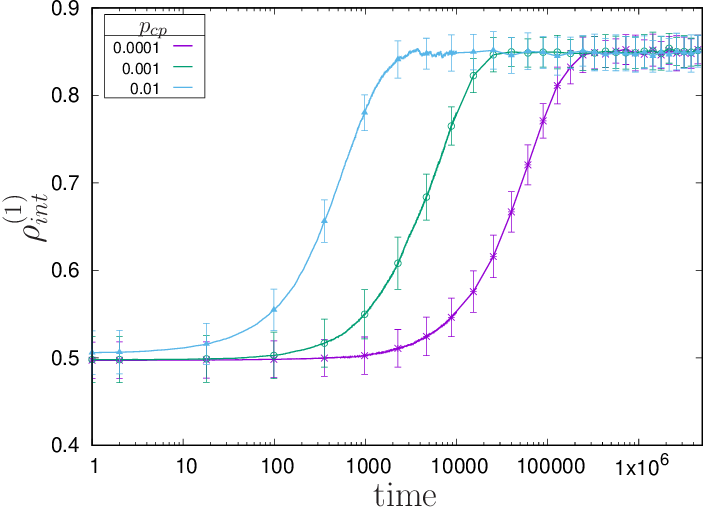}
\caption{Pluralistic ignorance level. The top panel shows the fraction of pluralistic ignorance as a function of the imitation probability ($p_{cp}$). The inset shows that even for $p_{cp}$ values as small as $0.0001$, the pluralistic ignorance fraction is still $\simeq0.85$. However, as $p_{cp}$ decreases, the time necessary for the system to reach an equilibrium increases, as can be seen in the bottom panel. The bars in the last panel stand for the standard deviation associated with $\rho_{int}^{(1)}$, while the lines are the average values of $\rho_{int}^{(1)}$.  Here $\gamma=0.9$ and $p_{int}=0.01$.}
\label{fig3}
\end{figure}

For the external state $1$ to propagate, not only does the imitation probability have to be lower than a threshold value, but also the influence factor $\gamma$ has to be significant. After the first individual externalizes the state $1$, the probability that the subsequent individual imitates this behavior is equal to ${(1/G)^{1-\gamma}}$. This probability increases with $\gamma$. Thus, a high value of the influence factor is required to provide a minimal chance for the external state $1$ to be imitated. If $\gamma$ is low, the population is absorbed in the configuration in which everyone is in the state $(0,0)$. 
The discontinuity of $\rho_{int}^{(1)}$ at $p_{cp}=0$ shown in Fig.~\ref{fig3} can be understood in terms of the effect of the first individual that externalizes the state $1$. First, if $p_{cp}=0$, individuals always externalize: those starting in $(1,0)$ end in $(1,1)$, and those in $(0,0)$ remain in $(0,0)$. If $p_{cp}$ is very small, there is nearly no imitation and individuals almost always externalize their internal state. Consequently, $(1,0)$ individuals almost always become $(1,1)$, and many individuals show $1$ as external states that can be imitated with probability at least ${(1/G)^{1-\gamma}}$, which is close to one for high $\gamma$. Thus, even if $p_{cp}$ is small, the few $(0,0)$ individuals that imitate will very likely adopt the external state $1$, becoming $(0,1)$, and with probability $p_{int}$, internalize $1$ to become $(1,1)$. This increases the fraction of individuals with $s_i=1$. The only way the fraction of individuals with $s_i=0$ can increase is if $(1,0)$ individuals do not imitate any external state $1$ and internalize the $0$. However, the probability that an individual does not imitate an external state equal to $1$ is at most equal to $(1-(1/G)^{1-\gamma})$, which is close to zero if $\gamma$ is sufficiently high. Thus, even if $p_{cp}$ is nearly zero, the fraction of individuals with $s_i=1$ increases more than those with $s_i=0$. The inset of Fig.~\ref{fig3} shows the pluralistic ignorance level for very small values of $p_{cp}$, which stays around $0.85$. As expected, the time needed to reach the stationary values increases as $p_{cp}$ becomes closer to zero (see the bottom panel of Fig.~\ref{fig3}).

\begin{figure}[h]
\includegraphics[width=\linewidth]{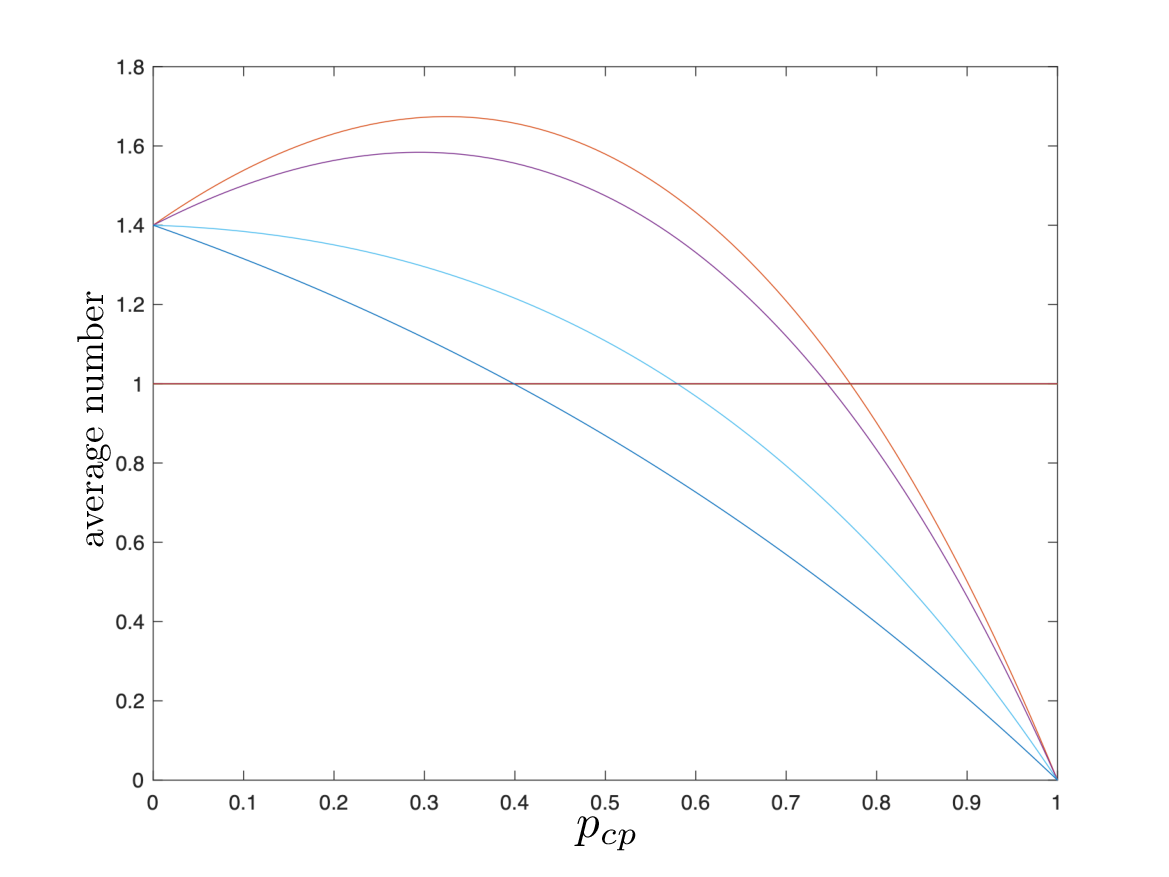}
\caption{Average number of individuals displaying external state equal to $1$ for a fraction of individuals with internal opinion $0$ equal to $0.8$. From top to bottom: $\gamma=0.99,0.9,0.5,0.0$. The horizontal line is the baseline. Here, we set $G=7$.}
\label{fig4}
\end{figure}

To understand the role played by the first individual that externalizes the state $1$, we can use a mean-field approach to calculate the average number of individuals displaying $1$ as their external state. First, let us define two auxiliary functions:
\begin{eqnarray}
    R(k)&=&p_{cp}\left(\frac{k}{G}\right)^{1-\gamma}+(1-p_{cp})(1-x_0) \mbox{ and } \\ 
    Q(k)&=&p_{cp}\left(1-\left(\frac{k}{G}\right)^{1-\gamma}\right)+(1-p_{cp})x_0,\\\nonumber 
\end{eqnarray}
where $k$ is the number of individuals in the group that have already displayed the external state $1$ and $x_0$ is the fraction of individuals with internal state equal to 0 in the population (notice that $1-x_0$ is the fraction with internal states equal to 1). Thus, $R(k)$ is the probability that a new external state equal to $1$ appears, whereas $Q(k)$ is the probability that an external state equal to $0$ is maintained if there are $k$ individuals already displaying an external state equal to $1$.
The probability $P_1(1)$ that the first individual in the group displays an external state equal to $1$ is given by $R(0)$ and the probability $P_1(0)$ that the first individual displays an external state equal to $0$ is given by $Q(0)$. After the display of the external state by the second individual, we have four potential outcomes: $11$, $10$, $01$, and $00$. Here, the first digit (on the left) denotes the external state of the first individual to exhibit the external state, while the second digit represents that of the second individual. Each of these outcomes occurs with probabilities given by ${P_2(1,1)=R(0)P(1)}$, ${P_2(1,0)=R(0)Q(1)}$, ${P_2(0,1)=Q(0)R(0)}$, and ${P_2(0,0)=Q(0)Q(0)}$,  respectively. Thus, if there are only two individuals in the group, the average number of individuals displaying an external state equal to $1$ is given by
\begin{equation}
    \bar{q}_2=2\cdot P_2(1,1)+1\cdot[P_2(1,0)+P_2(0,1)].
\end{equation}
Let us generalize these expressions. If $s_1$ is the state of the first individual that displays the external state, then the probability of the outcome $s_1$ is given by
\begin{equation}
    P_1(s_1)=s_1R(0)+(1-s_1)Q(0).
\end{equation}
The probability of the outcome $s_1s_2$, where $s_2$ is the external state of the second individual displaying it, is given by
\begin{equation}
    P_2(s_1,s_2)=P_1(s_1)[s_2R(s_1)+(1-s_2)Q(s_1)].
\end{equation}
Thus, after all $G$ individuals have the opportunity to display their external states, the probability of a specific outcome $s_1 s_2\ldots s_G$ is given by
\begin{eqnarray*}
P_G(s_1,\ldots,s_G)&=&P_{G-1}(s_1,\ldots,s_{G-1})\left[s_GR\left(\sum_{i-1}^{G-1}s_i\right) \right. \\ \nonumber
&&\left. +(1-s_G)Q\left(\sum_{i-1}^{G-1}s_i\right)\right]  
\end{eqnarray*}
and the average number of individuals displaying an external state equal to $1$ is given by
\begin{equation}
    \bar{q}_G=\sum_{s_1,\ldots,s_G}\left(\sum_{i=1}^Gs_i\right)P_G(s_1,\ldots,s_G).
\end{equation}
Figure~\ref{fig4} shows the average number of individuals displaying the external state $1$ for $x_0=0.8$. As we increase the value of $p_{cp}$, the average number of individuals displaying the external state $1$ initially increases, reaching a peak, and then starts to decrease, falling below a count of one. This peak occurs only when $\gamma$ is sufficiently high, underscoring the need to stimulate imitation to sustain the external presence of the state $1$. Furthermore, the average count drops below one when $p_{cp}$ approaches $0.8$, which represents the threshold at which the internal state 0 dominates the population.

\begin{figure}[h]
\includegraphics[width=\linewidth]{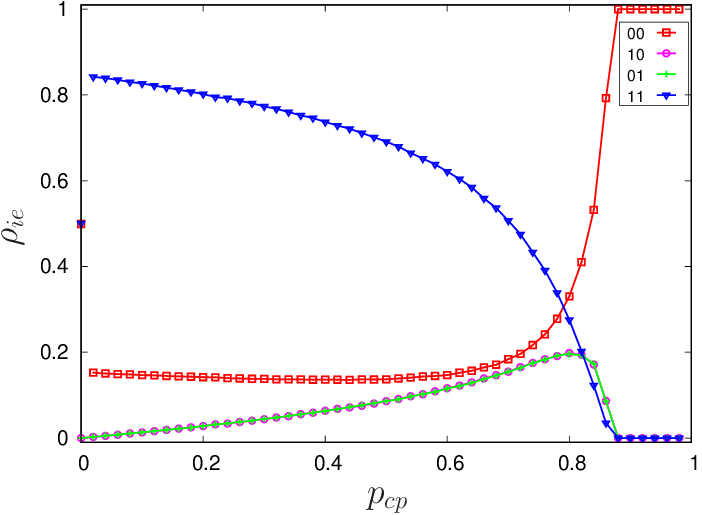}
\caption{Fractions $\rho_{ie}$ of each type $(s_i,s_e)$, at the end of the group interactions. The internalization probability and the influence factor are $p_{int}=0.01$ and $\gamma=0.9$, respectively.}
\label{fig5}
\end{figure}

Individuals break through the barrier of social pressure and overcome pluralistic ignorance when those with an internal state equal to $1$ are capable of externalizing their internal state. To observe this phenomenon, we quantify the fraction of each configuration $(s_i,s_e)$ after all individuals have had the opportunity to participate in a group and undergo the externalization, imitation, or internalization processes.  Figure~\ref{fig5} shows the fraction of individuals in each state after the population reaches the stationary state. If $p_{cp}$ is not too high, many $(1,0)$ individuals manage to change to $(1,1)$. However, although these $(1,1)$ individuals temporarily overcome pluralistic ignorance, they will continue to experience it because they start in the state $(1,0)$ in the next group interaction. As $p_{cp}$ increases, the fraction of individuals who manage to reach the $(1,1)$ state decreases until all individuals become in the state $(0,0)$.  Interestingly, the fraction of individuals experiencing conflict between internal and external states at the end of the group interaction increases with $p_{cp}$, reaches a peak near $0.8$, and then drops to zero. The reason for this non-monotonic behavior is simple. For low $p_{cp}$, the externalization is very frequent and most individuals end up at $(0,0)$ or $(1,1)$. As $p_{cp}$ increases, there is more imitation, and individuals in configurations $(0,1)$ or $(1,0)$ emerge. If $p_{cp}$ is too high, the external state equal to $1$ is not displayed by anyone, and, due to internalization, everyone eventually internalizes the external state $0$. Therefore, there must be a maximum at some intermediate value of $p_{cp}$, which happens to be close to $0.8$.

The internalization probability is crucial for the modification of internal states.  Our results are qualitatively unaffected by $p_{int}$. As $p_{int}$ increases, the fraction of dissonant states (states $(0,1)$ and $(1,0)$) at the end of group interactions is reduced because individuals internalize the external state, becoming harmonized more often. This mechanism can slightly change the fractions of $(1,0)$ and $(0,1)$ during the transient regime but does not affect the qualitative results. The most pronounced effect is the decrease in the $p_{cp}$ above which the $(0,0)$ state dominates the population.

As a final observation, it is worth considering why individuals, having overcome pluralistic ignorance (as evidenced by many ending up at $(1,1)$ for $p_{cp}$ not too high), do not retain that information and immediately begin the next interaction displaying $s_e=1$. This could be the case if the external state were merely a visible, permanent feature as in traditional opinion models. However, in our model, the root of pluralistic ignorance lies precisely in the initial absence of behavioral expression by everyone at the onset of interactions, which is interpreted as a norm enforced by social pressure.

\subsection{Internalization with memory}

How intense must an external state be experienced to enable its internalization, thereby transforming it into a permanent aspect of an individual's being?  It is worth considering that individuals might not internalize the current external state, but rather the most common external state. To investigate this point, we added memory to the model. Each individual stores the last $T_M$ external states adopted and internalizes the most frequent one instead of the latest one.

\begin{figure}[h]
\includegraphics[width=0.9\linewidth]{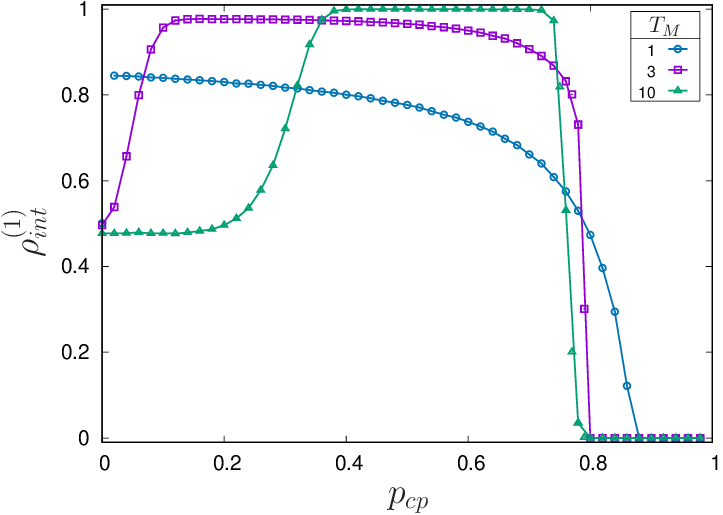}
\includegraphics[width=0.9\linewidth]{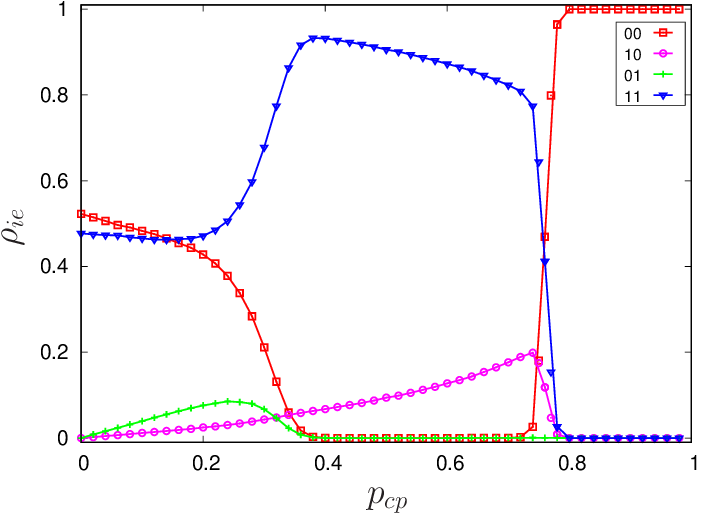}
\caption{The top figure shows the pluralistic ignorance level as a function of the imitation probability $p_{cp}$ for different memory lengths. The bottom figure shows the frequencies of $(s_i,s_e)$ at the end of the group interactions as functions of the imitation probability $p_{cp}$ for a memory length of $T_M=10$. Here the internalization probability is $p_{int}=0.01$ and the influence factor is $\gamma=0.9$.}
\label{fig6}
\end{figure}

Figure~\ref{fig6} shows the level of pluralistic ignorance for different memory lengths. For all memory lengths, the level of pluralistic ignorance increases as $p_{cp}$ increases, reaches a high-value plateau, and goes to zero at a high $p_{cp}$ value, mirroring the behavior observed previously for $T_M=1$. The main difference from the case $T_M=1$ is that pluralistic ignorance is high only at intermediate values of $p_{cp}$. The main reason for this behavior is that for $T_M>1$ individuals can internalize a state that is different from the state displayed in the last group interaction. More specifically, the $(0,0)$ individuals that become $(0,1)$ must internalize state 1 to become $(1,1)$. However, because the probability of imitation is low, the $(0,0)$ individuals will remain as $(0,0)$ most of the time, and the memory will be full of $0$'s.

\begin{figure*}[h]
\includegraphics[width=0.52\linewidth]{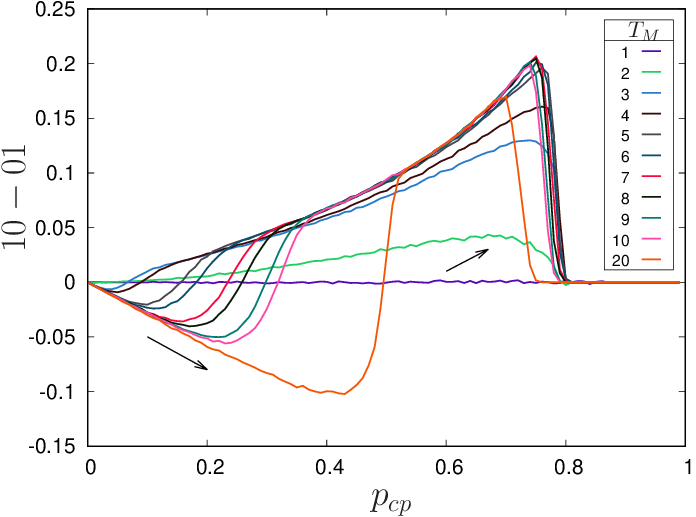}
\includegraphics[width=0.52\linewidth]{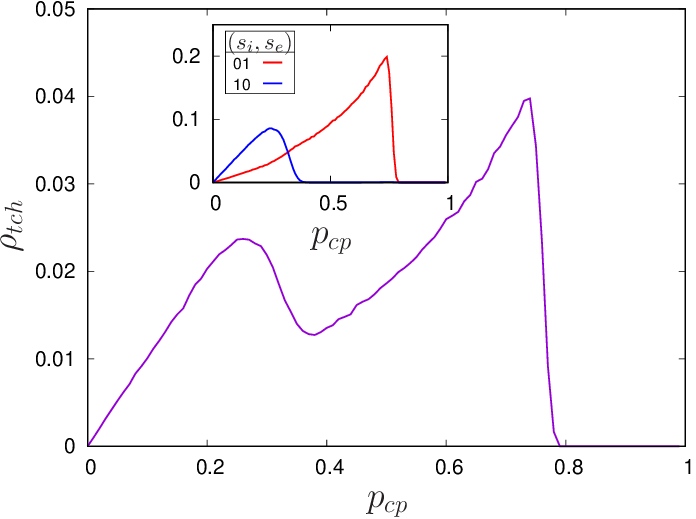}
\caption{Dissonance. The left panel shows the difference between the frequencies of dissonant states $(0,1)$ and $(1,0)$ as a function of the imitation probability $p_{cp}$ for different memory lengths. The right panel shows the frequency of internalization events that involve internalizing a state different from the current external one. The inset in the right panel shows the levels of dissonant states for the same system. Here, $p_{int}=0.01$ and $\gamma=0.9$.}
\label{fig7}
\end{figure*}

The effect of memory length is more intricate. For $T_M=1$, the fraction of $(1,0)$ and $(0,1)$ at the end of the group interactions are equal. However, as shown in Fig.~\ref{fig6}, for $T_M>1$, these fractions are no longer the same. Figure~\ref{fig7} shows in detail this asymmetry introduced by memory. As can be seen in the first panel of Fig.~\ref{fig7}, the difference in the frequencies of states $(0,1)$ and $(1,0)$ becomes greater as the memory length, $T_M$, increases. There are more $(0,1)$ states for low values of $p_{cp}$, the opposite happens for high $p_{cp}$. Not only does this difference increase with $T_M$, but also the value of $p_{cp}$ above which $(1,0)$ becomes higher than $(0,1)$ increases. The main reason for such behavior is indicated in the second panel of Fig.~\ref{fig7}, where the frequency $\rho_{tch}$ of internalization events in which the internalized state is different from the last external state is shown. Transitions from $(1,1)$ to $(0,1)$, for example, are now possible through internalization (when $0$ was the most frequently adopted external state in the last $T_M$ transitions). Similarly, transitions from $(0,0)$ to $(1,0)$ can now happen if $1$ was the most frequently adopted external state in the last $T_M$ transitions. Additionally, states $(0,1)$ can internalize $0$ and stay as $(0,1)$, as well as state $(1,0)$ can internalize $1$ and stay as $(1,0)$. 

For low values of $p_{cp}$, $0$ is the most abundant external state (see Fig.~\ref{fig6}). Thus, the internalization of $0$ is more common than that of $1$ and both states $(0,0)$ and $(0,1)$ tend to internalize $0$ and stay as $(0,0)$ and $(0,1)$, respectively. The introduction of memory in this case increases both the frequency of $(0,0)$ and $(0,1)$ when the value of $p_{cp}$ is low. For high values of $p_{cp}$, on the other side, states $(0,1)$ and $(0,0)$ practically disappear, since now $1$ is the most frequent external state, being internalized more frequently. Thus, states $(1,0)$ and $(1,1)$ become more abundant for higher values of $p_{cp}$. Notice that the frequency $\rho_{tch}$ in the second panel of Fig.~\ref{fig7} has two maximum values. The first one, which happens for a low value of $p_{cp}$, is due to the internalization of $0$ by individuals with external state $1$. The second maximum happens for a high value of $p_{cp}$ and is caused by the internalization of $1$ by individuals with external state $0$.

\subsection{Square lattice}

Finally, we tested the effect of adding spatial correlations to the model. We set each individual at a different site in a square lattice of size $22\times22$ with periodic boundary conditions. At each time step, one individual is randomly chosen to be the center of a group of size $G=21$. The groups are formed considering the Euclidean distance from a central site, with the distance between the two first neighbors set to one. For a group of 21, we add all neighbors that are located up to $\sqrt{5}$ units from the central site (Moore neighborhood of range 2 without corners).

The pluralistic ignorance level is very similar to the well-mixed case shown in~Fig.~\ref{fig2}. This is expected since the external states are set to zero at the beginning of each group interaction. Thus, there are not many opportunities for the formation of external state clusters, as in opinion dynamic models. 

\begin{figure*}[h]
\includegraphics[width=0.55\linewidth, valign=t]{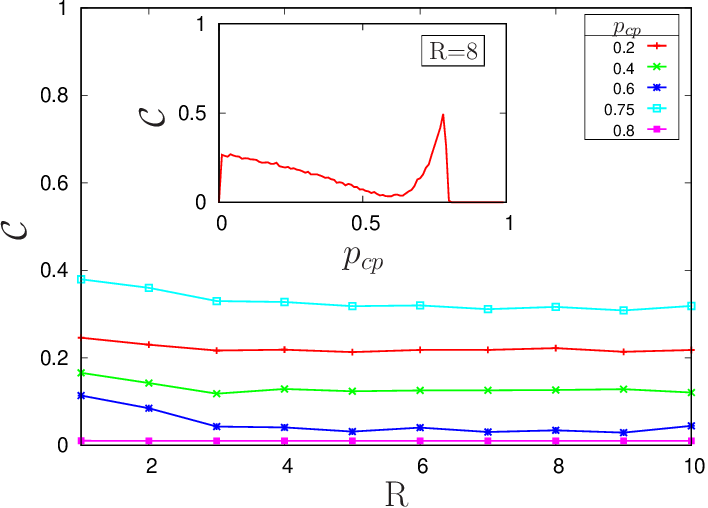}
\includegraphics[width=0.4\linewidth, valign=t]{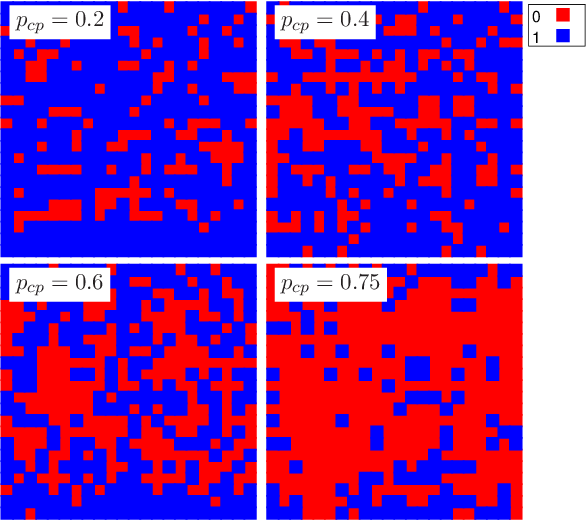}
\caption{Spatial correlation and configurations for the internal opinions in the square lattice. On the left, the multiple lines show the behavior of the spatial correlation as a function of the distance for different values of the imitation probability $p_{cp}$. The inset shows the spatial correlation as a function of $p_{cp}$ for $R=8$. Notice that, since the system is not that much larger than the group size, there is almost no variation in the spatial correlation as the distance changes. At the right, four different spatial configurations are shown, each one related to a different $p_{cp}$ value. Here, $\gamma=0.9$, $p_{int=0.01}$, $T_M=1$ and the number of neighbors for each agent is 20.}
\label{fig8}
\end{figure*}

Figure~\ref{fig8} shows the spatial correlations of the internal state of the individuals in the square lattice. Spatial correlations are almost the same for different distances $R$. However, as the imitation probability $p_{cp}$ changes, the average correlation varies. The inset in the first panel shows that the spatial correlation for $R=8$ has two maximum values: one for a small value of $p_{cp}$ and another for a high $p_{cp}$, just before the correlation reaches zero. Notice that the results for $R=8$ are representative of the other values of $R$ since the correlation is almost the same for all $R$. For $p_{cp}=0.2$, the external state $1$ is more abundant than $0$. As $p_{cp}$ increases, the difference in the amount of external states $1$ and $0$ decreases, until the state $0$ starts to be more popular than 1. These two extremes, one with the state $0$ being much more frequent and the other where $1$ is the most popular state, explain the maximum correlation values observed in the first panel.

\section{\label{conclusion}Discussion and Conclusions}

In this work, we introduced a simple model, similar to external and concealed opinion models, to investigate the emergence of pluralistic ignorance.  Our results show that in the absence of memory, pluralistic ignorance occurs for a wide range of values of imitation probability if the influence factor is strong. However, if there is memory, pluralistic ignorance emerges at high levels only for moderate values of imitation probability. Interestingly, if the imitation probability is not too high, at the end of the group interactions the population reaches a stationary state where most individuals are able to overcome pluralistic ignorance. However, individuals return to pluralistic ignorance at the beginning of the next group interaction. Most interestingly, the fraction of individuals that end the group interaction phase with conflicting internal and external states reaches a peak at an intermediate value of the imitation probability. If the probability of imitation is too high, the individuals eventually internalize the inaction and harmonize themselves in the state $(0,0)$.

Our model highlights that pluralistic ignorance can occur every time a group of strangers meet. Because $s_e=0$ is always the initial external state of everyone in the group, we used the fraction of individuals with $s_i=1$ as a proxy of pluralistic ignorance in the population. The fact that many in the group can overcome the pluralistic ignorance at the end of group interactions, ending in the state $s_i=1$ and $s_e=1$, does not mean that they will readily express their internal state in the next group interaction because at the beginning there is always the large group silence exerting social pressure.

Pluralistic ignorance arises from the complex interplay between individuals' hidden thoughts and feelings and the externally visible behaviors that are widely recognized within a group. These visible behaviors are often perceived as normative; more specifically, they represent descriptive norms~\cite{bicchieri2005grammar,Cialdini1991AFT}. Descriptive norms encompass the perception or understanding of what behaviors are typically practiced or approved within a specific group or society. They reflect observations or beliefs about common actions in a given situation, influencing individuals' behavior as they seek to conform to perceived societal standards. The descriptive norm is enforced in our model by the imitation rule. Using the classroom example, students initially see that their partners are not raising their hands and believe that because most are acting in this way, this is the dominant behavior in the group.

The phenomenon of pluralistic ignorance can be the consequence of behavioral responses characterized by inaction if this is interpreted as a cue for the group's normative behavior. This notion is epitomized by Plato's famous saying, ``silence gives consent'', highlighting that the absence of behavior can leave marks in the world. In situations where the internal states of the others are readily observable,  pluralistic ignorance cannot be present. For example, if there was a neurological mechanism that caused individuals to consistently frown when in doubt, students would immediately recognize their peers' lack of comprehension of a subject. We should contrast this mechanism to the alcohol-drinking case where the normative behavior is the action of drinking. One might be tempted to say that we could use our model to understand drinking behavior by interpreting $s_e=0$ as drinking behavior and $s_1=1$ as an internal attitude against abuse of alcohol consumption. However, at the very beginning of a group interaction (in this example, it would be a campus party), the real behavior would be no drinking at all, which does not match the social norm saying that excessive drinking is cool. To model this specific instance of pluralistic ignorance, we would need to model the social norm explicitly.

There is a similar phenomenon called the bystander effect, or bystander apathy, which is closely linked to pluralistic ignorance and can also be understood using our results. The bystander effect is a social phenomenon in which individuals are less likely to offer help or intervene in an emergency situation when other people are present. This effect suggests that the presence of others can inhibit individual responsibility and action, leading to the diffusion of responsibility among bystanders~\cite{Rendsvig2014-RENPII}. This phenomenon is often associated with pluralistic ignorance. At the beginning of the group interaction, people may want to help. However, as they see that nobody is helping, they start to believe that they are the only ones who think that. This is an example of how failing to overcome pluralistic ignorance can lead to profoundly negative consequences.

Internalization is often defined as the incorporation of a norm into one's personality in such a way that the agent acts accordingly even in the absence of social pressure~\cite{campbell1964}. Here, we do not model the norm explicitly. Instead, the norm is inferred from the initial state of the group and is expressed in the tendency to imitate others. Using the classroom example, one would say that ``because everyone is showing 0, this must be the group standard''. More specifically, we adopt the concept of internalization as defined by Bem's self-perception theory, which states that individuals subconsciously adjust their internal beliefs to align with their stated opinions~\cite{bem1967}.

Our model shares similarities with both the Bass diffusion model~\cite{Bass:1969} and Granovetter threshold models of collective behavior~\cite{aGranovetter1978}. In these models, a fundamental challenge lies in elucidating the emergence of novel behaviors within the context of collective action. The Bass model posits that certain individuals adopt a new idea independently of their social circle, subsequently influencing others through the evolving dynamics of the social environment. As adoption spreads, the impact on later adopters amplifies, driven by the increasing number of prior adopters. Similarly, the Granovetter model emphasizes individual thresholds, wherein the decision to adopt a behavior hinges on the actions of a certain number of peers. Innovators in the Bass model and individuals with low thresholds in the Granovetter model serve as catalysts, sparking collective adherence to the action. In our model of pluralistic ignorance, initial externalizations function as catalysts, incentivizing subsequent individuals to imitate the behavior. Despite this similarity, we stress that both Bass and Granovetter's models do not take into account the dual layer of internal and external states.

In summary, pluralistic ignorance is a group phenomenon with potentially adverse consequences. Instances such as public concern about climate change, racial segregation, and exaggerated alcohol consumption exemplify scenarios in which individuals' perceptions of others diverge from reality, resulting in detrimental outcomes for all involved. Whenever the absence of a behavioral expression that is under normative constraint matches the normative status of the activity (for example, initially not raising the hands matches the normative behavior of not asking questions),  pluralistic ignorance can emerge as a consequence of social pressure to conform.

\section*{Credit authorship contribution statement}
\textbf{A.F.Lütz:} conceptualization, data curation, formal analysis, investigation, methodology, software, validation, visualization, writing – original draft, writing – review \& editing. \textbf{L.Wardil:} conceptualization, data curation, formal analysis, investigation, methodology, software, validation, visualization, supervision, project administration, writing – original draft, writing – review \& editing.

\section*{Acknowledgments}
 This work was supported by the Brazilian Research Agency CAPES (proc. 88887.463878/2019-00) and the Minas Gerais State Agency for Research and Development FAPEMIG (proc. APQ-00694-18).   

\section*{Declaration of competing interest}
The authors declare that they have no known competing financial interests or personal relationships that could have appeared to influence the work reported in this paper.


\end{document}